\documentclass[12pt,preprint]{aastex}

\newcommand{\kms}{km~s$^{-1}$}

\newcommand{\ta}{G23.01--0.41}
\newcommand{\tb}{G23.44--0.18}
\def\masy {mas~y$^{-1}$}

\def\etal {et al.~}

\def\ie   {i.e.~}

\def\Vlsr {\ifmmode {V_{\rm LSR}} \else {$V_{\rm LSR}$} \fi}
\def\Ro   {\ifmmode {R_0} \else {$R_0$} \fi}
\def\To   {\ifmmode {\Theta_0} \else {$\Theta_0$} \fi}

\slugcomment{Version: 2008 November 05}

\shorttitle{Distance to \ta\ and  \tb}
\shortauthors{Brunthaler \etal}

\begin{document}

\title{Trigonometric Parallaxes of Massive Star Forming Regions: V. 
       \ta\ and \tb}

\author{A. Brunthaler\altaffilmark{1},M. J. Reid\altaffilmark{2}, 
        K. M. Menten\altaffilmark{1},X. W. Zheng\altaffilmark{3},
        L. Moscadelli\altaffilmark{4}, and Y. Xu\altaffilmark{1,5}}

\altaffiltext{1}{Max-Planck-Institut f\"ur Radioastronomie,
   Auf dem H\"ugel 69, 53121 Bonn, Germany}
\altaffiltext{2}{Harvard-Smithsonian Center for
   Astrophysics, 60 Garden Street, Cambridge, MA 02138, USA}
\altaffiltext{3}{Department of Astronomy, Nanjing University,
   Nanjing 210093, China}
\altaffiltext{4}{INAF, Osservatorio Astrofisico di Arcetri, Largo E. Fermi 5, 50125 Firenze, Italy}
\altaffiltext{5}{Purple Mountain Observatory, Chinese Academy of
Sciences, Nanjing 210008, China}

\begin{abstract}
We report trigonometric parallaxes for the massive star-forming regions
\ta\ and \tb, corresponding to distances of $4.59^{+0.38}_{-0.33}$~kpc
and $5.88^{+1.37}_{-0.93}$~kpc, respectively. The distance to \ta\ is 
smaller than its near kinematic distance assuming a standard model of the 
Milky Way and less than half of its far kinematic distance, which has usually 
been assumed. This places it in the Crux-Scutum spiral arm.  The distance 
to \tb\ is close to its near kinematic distance and most likely places it 
in the Norma spiral arm near the end of the Galactic bar. Combining the 
distance and proper motions with observed radial velocities gives the 
location and full space motion of the star forming regions.
We find large deviations from circular Galactic orbits
for these sources: both sources show peculiar motions 
of 20 to 30 \kms\ counter to Galactic rotation and toward the 
Galactic center.  These large peculiar motions might be the
result of gravitational perturbations from the Galactic bar.
\end{abstract}


\section{Introduction}

As part of a large program to determine Galactic structure
by measuring trigonometric parallaxes and proper motions of 
newly formed stars, we are conducting multi-epoch VLBA 
observations of methanol masers associated with high-mass 
star forming regions.  The motivation and background
information regarding our program are discussed in 
\citet{ReidMentenBrunthaler2008a}, hereafter called Paper I.   

In this paper, we report observations of 12 GHz methanol masers
toward two regions of high-mass star formation:  \ta\ and \tb.  
These sources are toward Galactic longitude $23^\circ$ and both 
have large Local Standard of Rest (LSR) velocities, suggesting they 
are at great distances.  Since high-mass star forming regions
define spiral arms, accurate parallaxes for these sources 
should help locate the inner spiral arms of the Galaxy.  
Additionally, the Galaxy is thought to possess a central bar
that terminates near the longitude of these star forming regions.
Thus, by locating the masers and measuring their 3-dimensional
space motions, we hope to learn about the effects of the bar
potential.

\section{Observations and Calibration}

We used the NRAO
\footnote{The National Radio Astronomy Observatory is a facility of the
National Science Foundation operated under cooperative agreement by
Associated Universities, Inc.}
Very Long Baseline Array (VLBA) to observe the 12 GHz methanol 
masers in the two star forming regions \ta\ and \tb.
Paper I describes the general observational strategy and data calibation
procedures.  In this section, we describe only those aspects of the
observations that are specific to these two sources.

The observations conducted under program BR100F were scheduled
on 2005 October 28, 2006 March 22 and September 29, and 2007 March 18.  
These dates were chosen to sample the peaks of the parallax 
signature in Right Ascension only, as the amplitude of the signature 
in Declination is considerably smaller.  The total observing time for 
each epoch was 10 hours.

\begin{deluxetable}{lllllcl}
\tablecolumns{7} 
\tablewidth{0pc} 
\tablecaption{Source Information}
\tablehead {
  \colhead{Source} & \colhead{R.A. (J2000)} &  \colhead{Dec. (J2000)} &
  \colhead{$\phi$}& \colhead{P.A.} & \colhead{\Vlsr} &
  \colhead{Restoring Beam}
\\
  \colhead{}       & \colhead{(h~~m~~s)} &  \colhead{($^\circ$~~'~~'')} &
  \colhead{($^\circ$)}&\colhead{($^\circ$)} & \colhead{(\kms)} &
  \colhead{(mas, mas, deg)}
            }
\startdata
 G\,23.01 ......& 18 34 40.27700  & -09 00 38.4010 &  0  &...  & 81.5 & $2.7\times0.8$ @ $-14$  \\
 J1825-0737 ... & 18 25 37.60955  & -07 37 30.0128 &  2.6&-58       &  & $2.1\times0.7$ @ $-13$    \\
 J1832-1035 ... & 18 32 20.83656  & -10 35 11.2006 &  1.7&-20       & & $2.1\times0.7$ @ $-12$     \\
 J1833-0855 ... & 18 33 19.58200  & -08 55 27.2100 &  0.3&-75       &  & $2.1\times0.7$ @ $-14$    \\
\\
 G\,23.44 ......& 18 34 39.18690  & -08 31 25.4050 &0 &... & 97.6 & $3.3\times2.0$ @ $-1$\\
 J1825-0737 ... & 18 25 37.60955  & -07 37 30.0128 &  2.4&-68       & & $2.6\times1.2$ @ $-4$     \\
 J1832-1035 ... & 18 32 20.83656  & -10 35 11.2006 &  2.1&-15       &  & $2.7\times1.2$ @ $-2$    \\
 J1833-0855 ... & 18 33 19.58200  & -08 55 27.2100 &  0.5&-39       &  & $2.7\times1.2$ @ $-2$    \\
\enddata
\tablecomments {$\phi$ and P.A. are the separations and position angles 
(East of North) between maser and reference sources. 
The radial velocity of the masers and the size and shape of the
interferometer restoring beam are listed for the first epoch’s data.
               }
\end{deluxetable}


For background position references, we observed two sources from the 
VLBA Calibrator Survey, J1825-0737 \citep{VCS2} and  
J1832-1035 \citep{VCS4}, and one source from a targeted VLA calibrator 
search, J1833-0855 \citep{XuReidMenten2006}. 
We used the ICRF source J1800+3848 \citep{ICRF} as a fringe finder. 
For the phase-referencing observations, we used four adjacent frequency 
bands with 4 MHz bandwidth in both right and left circular 
polarization.  The second band contained the maser signals and
was centered at an LSR velocity of 75.0~\kms\ for \ta\ and 
103.0~\kms\ for \tb.

\begin{figure}
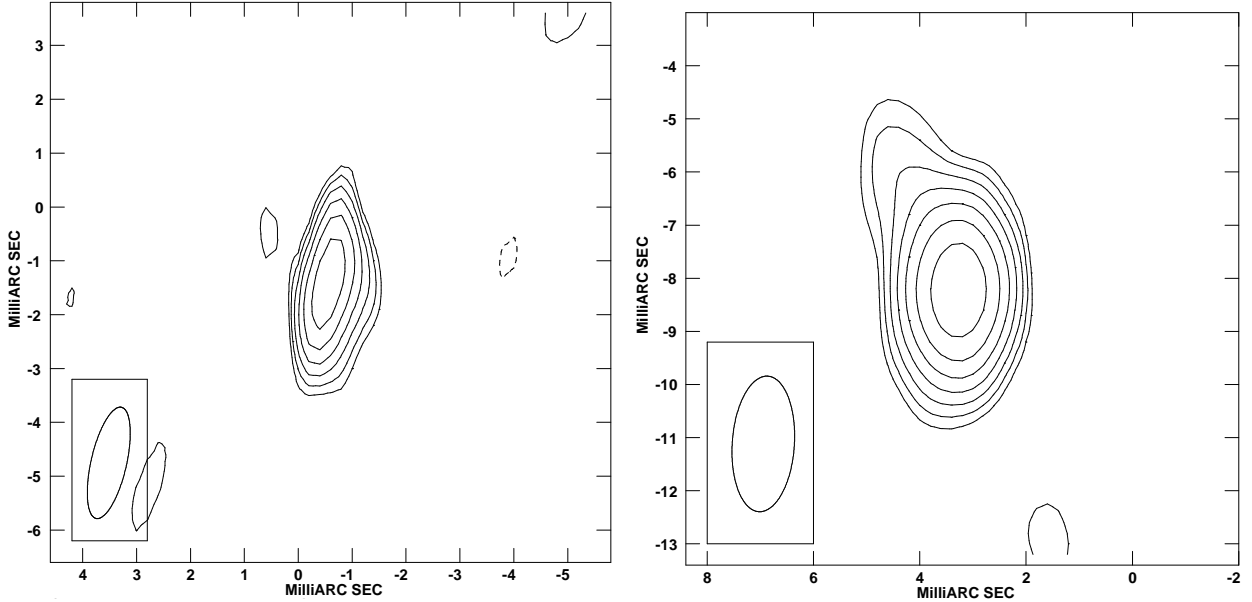

\resizebox{0.5\hsize}{!}{\includegraphics[bbllx=1.5cm,bburx=20.2cm,bblly=5.6cm,bbury=23.7cm,clip=,angle=0]{f1a.ps}}
\resizebox{0.5\hsize}{!}{\includegraphics[bbllx=1.5cm,bburx=20.2cm,bblly=5.8cm,bbury=23.6cm,clip=,angle=0]{f1b.ps}}



\caption{Phased-referenced images of J1825-0737 using \ta\,(left) and \tb\,(right) as reference source from the first epoch. The contours start at 12 mJy and increase with $\sqrt{2}$. Beam FWHM is indicated at lower left.}
\label{qso}
\end{figure}

We performed a ``manual phase-calibration'' using the data
from J1800+3848 to remove instrumental phase offsets among the 
four frequency bands.  We used a spectral channel with one bright 
and compact maser feature as the phase reference:
for \ta\ we used the channel at \Vlsr\ of 81.54 \kms\ and for \tb\  
the channel at 97.62 \kms.

\section{Results}

After calibration, we imaged the maser emission and the continuum sources 
using the AIPS task IMAGR.  While all three background sources were 
clearly detected, J1832-1035 and J1833-0855 were heavily resolved 
and could not be used for high precision astrometry.
Fig.~\ref{qso} shows for the first epoch the images of J1825-0737 
using \ta\,(left) and \tb\,(right) as the phase-reference. 
The small deviations from a point-like image are most likely caused by 
errors in the phase calibration due to short term fluctuations in the 
troposphere. Since the reference maser in \tb\ is resovled on the longest 
baselines, the data from the telescope on Mauna Kea could not be used for this
source. This explains the different beam sizes for the sources in Table 1 and
Fig.~\ref{qso}.

We estimated the positions of the masers and the background source 
by fitting elliptical Gaussian brightness distributions to the images. 
The positions of the maser spots relative to the background source were 
then modeled by the parallax sinusoid (including the effects of the 
ellipticity of the Earth's orbit) and a proper motion in each coordinate. 
Since systematic errors generally dominate over random noise, 
formal position errors are usually unrealistic small.  This results in 
relatively high $\chi^2$ per degree of freedom values of the parallax fit 
(typically values between 2 and 8).  The magnitude of the systematic errors 
are {\it a priori} unknown and can only be estimated from the goodness of 
the fits.  Hence, we assigned independent ``error floors'' to the Right 
Ascension and Declination positions and added them in quadrature to 
the formal position fitting errors.  The error floors were then 
adjusted until residual $\chi^2$ per degree of freedom values 
of unity were achived for each coordinate.

\subsection{\ta}
\begin{figure}
\resizebox{\hsize}{!}{\includegraphics[bbllx=1.9cm,bburx=19.6cm,bblly=3.9cm,bbury=25.4cm,clip=,angle=0]{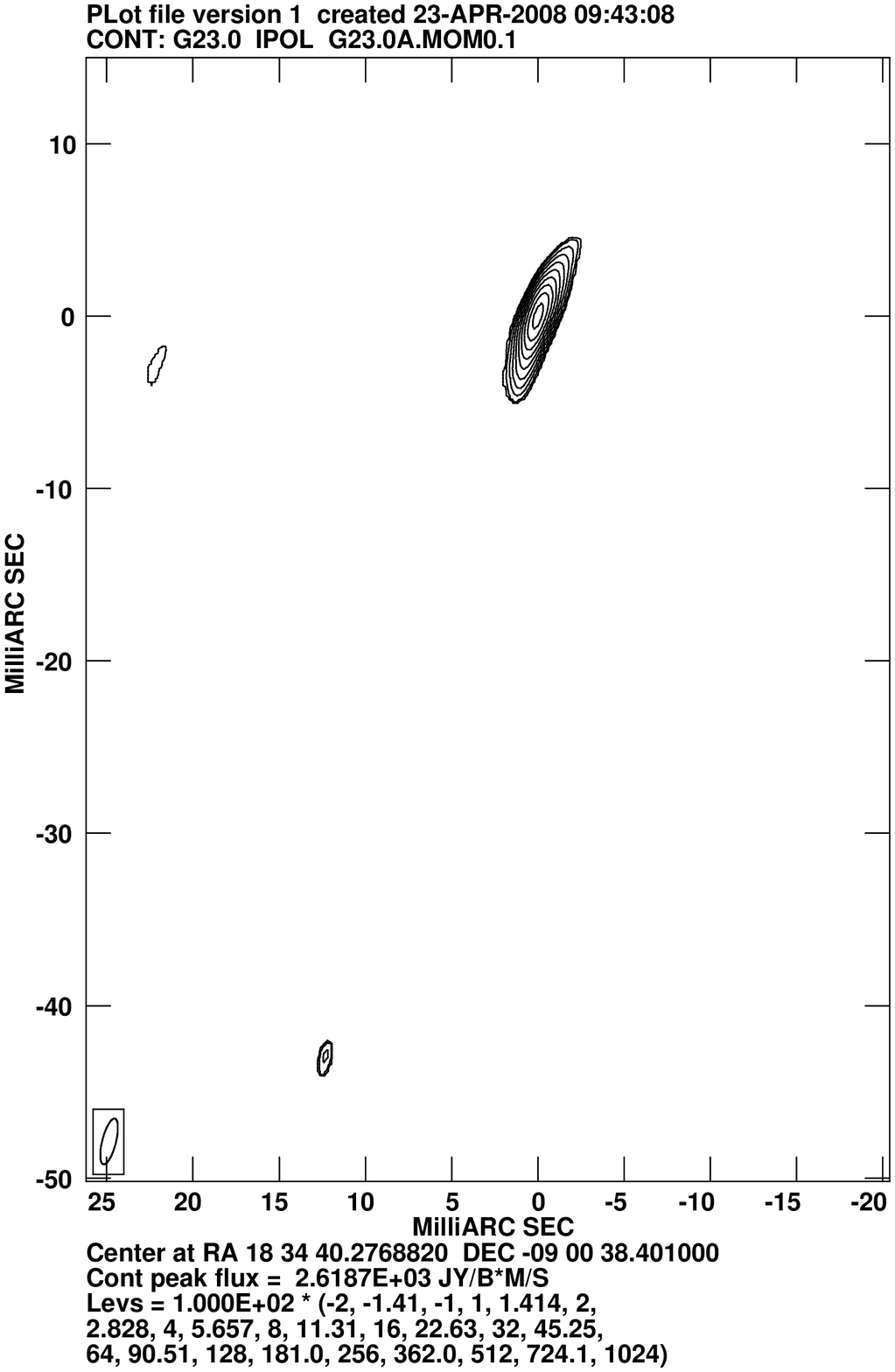}}
\caption{Velocity integrated maser emission of \ta\, in the first epoch. The contours start at 0.1 Jy km s$^{-1}$ and increase with~$\sqrt[]{2}$. Beam FWHM is indicated at lower left.}
\label{mom0}
\end{figure}

\begin{figure}
\resizebox{\hsize}{!}{\includegraphics[angle=-90]{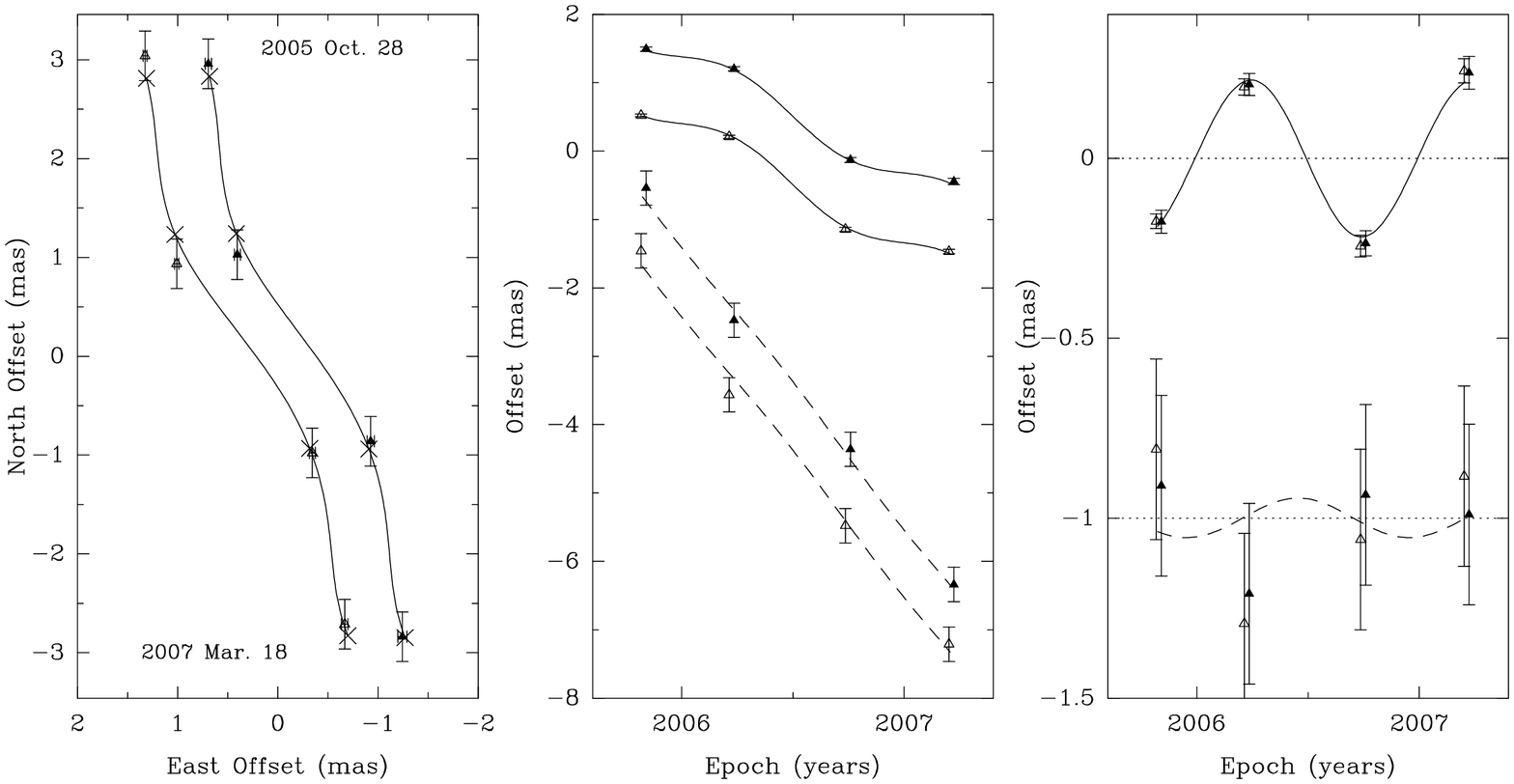}}
\caption{Results of the parallax fit for \ta. The different symbols represent 
the two maser spots. {\bf Left Panel:} Sky projected motion of the maser. The
crosses and the lines show the best-fit position offsets and the trajectrory,
respectively. {\bf Middle Panel:} The position offfsets of the masers along
the East and North direction versus time. The best-fit model in East and North
direction are shown as continuous and dashed lines, respectively. {\bf Right
  Panel:} Same as the middle panel but with fitted proper motions removed. The
North offset data have been shifted for clarity.}
\label{g23.0}
\end{figure}

The maser emission toward \ta\, consists of one bright and one weak spot 
separated by $\sim45$ mas (see Fig.~\ref{mom0}), both of which could be 
used for precision astrometry. The methanol maser position agrees within 0.3'' 
with the position of a recently detected H$_2$CO maser and the peak of the 
NH$_3$ (3,3) emission \citep{ArayaHofnerGoss2008}. Table~\ref{table:G0_ppm} 
and Fig.~\ref{g23.0} show the results of the parallax and 
proper motion fit for each maser spot. The fact that we obtain equally good 
fits for the strong and the weak maser spot shows that our accuracy is not
limited by the signal to noise ratio but by systematic errors. A combined fit, 
where the positions
of the two maser spots are fitted with one parallax but different proper 
motions yields a parallax of $0.218$ with a formal uncertainty of
$\pm 0.012$ mas. However, since there appear to be significant 
correlation among the different spots at each epoch (as evidenced by
a tendency for data to deviate from the fit in the same direction),
we take a conservative approach and adopt a parallax uncertainty
of $\pm 0.017$ (\ie, multipling the formal uncerainty by $\sqrt[]{2}$). 
Hence, we obtain a distance of $4.59^{+0.38}_{-0.33}$~kpc for \ta.

Our measured distance is smaller than the near kinematic distance of 5.0~kpc, 
assuming IAU values for the distance to the Galactic center of $R_0=8.5$ 
kpc and for the circular rotation speed of the LSR of $\Theta_0=220$~\kms. 
The measured distance is only $\sim 43\%$ of the far kinematic 
distance, which has been assumed in most previous studies of this region 
\citep{CodellaTestiCesaroni1997, FuruyaCesaroniTakahashi2008, 
ArayaHofnerGoss2008}. Thus, physical sizes and luminosities have been 
overestimated in these studies by factors of 2.4 and 5.5, respectively.
The near kinematic distance can be brought into better agreement with the
measured parallax, if one uses a Galactic rotation model with 
$\Theta_0/R_0=29.5$~\kms\ kpc$^{-1}$ \citep{ReidBrunthaler2004}.

The measured distance places \ta\ at a Galactocentric radius of 4.64 kpc.
The proper motions of both maser spots are consistent with each other, 
indicating that internal motions are small, as observed in many methanol maser 
sources \citep{MoscadelliMentenWalmsley2002}.  The average 
proper motion is $-1.72\pm 0.04$~\masy\ toward the East and 
$-4.12\pm 0.3$~ \masy\ toward the North, corresponding to $-37\pm3$~\kms\ and 
$-90\pm10$~\kms, respectively. 

\begin{deluxetable}{lllll}
\tablecolumns{5} \tablewidth{0pc}
\tablecaption{G23.01-0.41 Parallax \& Proper Motion Fits}
\tablehead {
  \colhead{Maser \Vlsr} & \colhead{Background} &
  \colhead{Parallax} & \colhead{$\mu_x$} &
  \colhead{$\mu_y$}
\\
  \colhead{(\kms)}      & \colhead{Source} &
  \colhead{(mas)} & \colhead{(\masy)} &
  \colhead{(\masy)}
            }
\startdata
 81.5& J1825-0737 &$0.217\pm0.019$ &$-1.74\pm0.04$ &$-4.11\pm0.3$ \\
 80.8& J1825-0737 &$0.221\pm0.020$ &$-1.70\pm0.04$ &$-4.13\pm0.3$ \\
\\
 combined &             &$0.218\pm0.017$ \\

\enddata
\tablecomments {Combined fit used a single parallax parameter
for both maser spots relative to the background source;
a single proper motion was fit for each maser spot relative
to the background source.}
\label{table:G0_ppm}
\end{deluxetable}

\subsection{\tb}

\begin{figure}
\resizebox{\hsize}{!}{\includegraphics[bbllx=3.2cm,bburx=17.0cm,bblly=1.4cm,bbury=26.3cm,clip=,angle=-90]{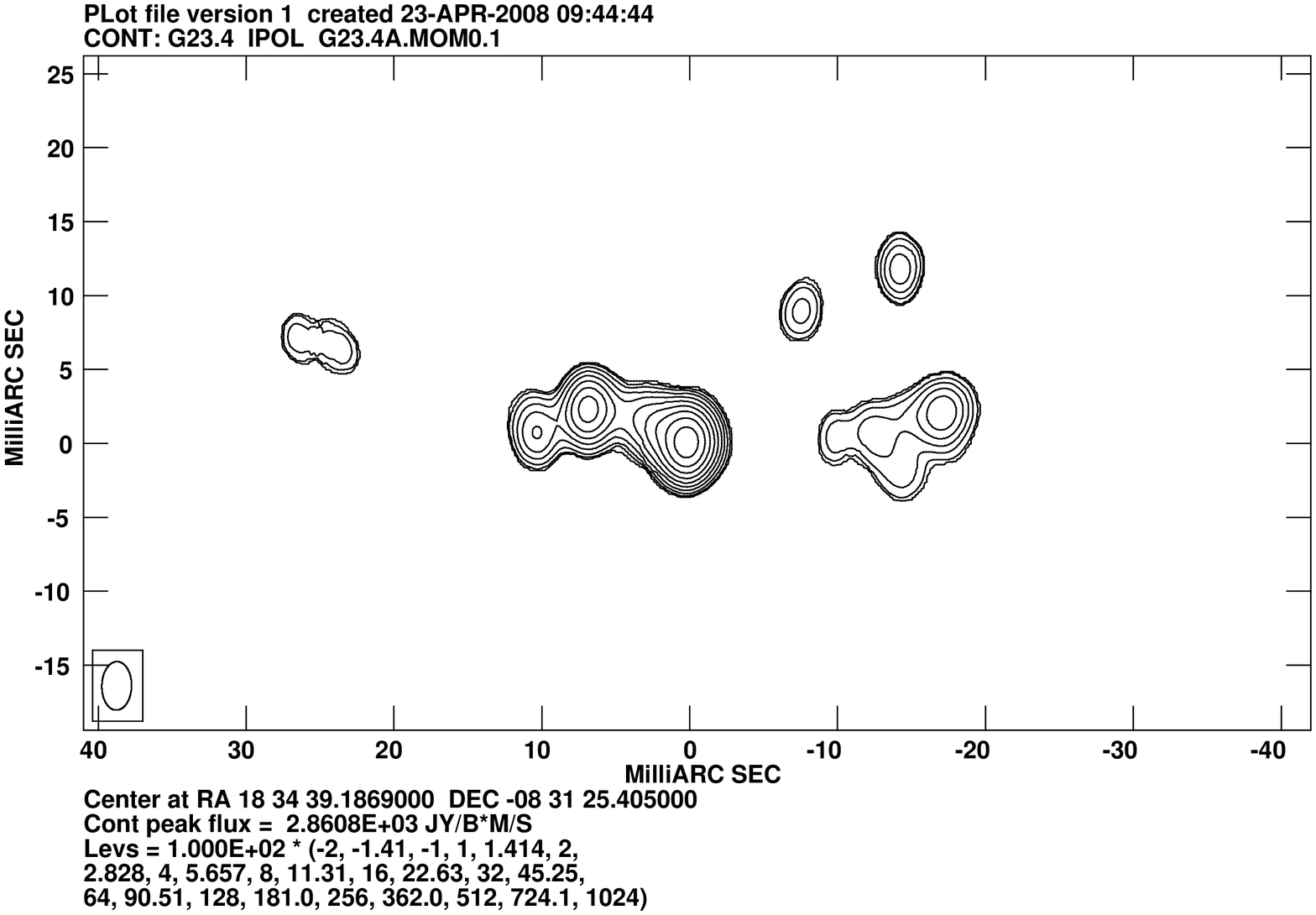}}
\caption{Velocity integrated maser emission of \tb\, in the first epoch. The contours start at 0.1 Jy km s$^{-1}$ and increase with~$\sqrt[]{2}$. Beam FWHM is indicated at lower left.}
\label{mom4}
\end{figure}

\begin{figure}
\resizebox{\hsize}{!}{\includegraphics[angle=-90]{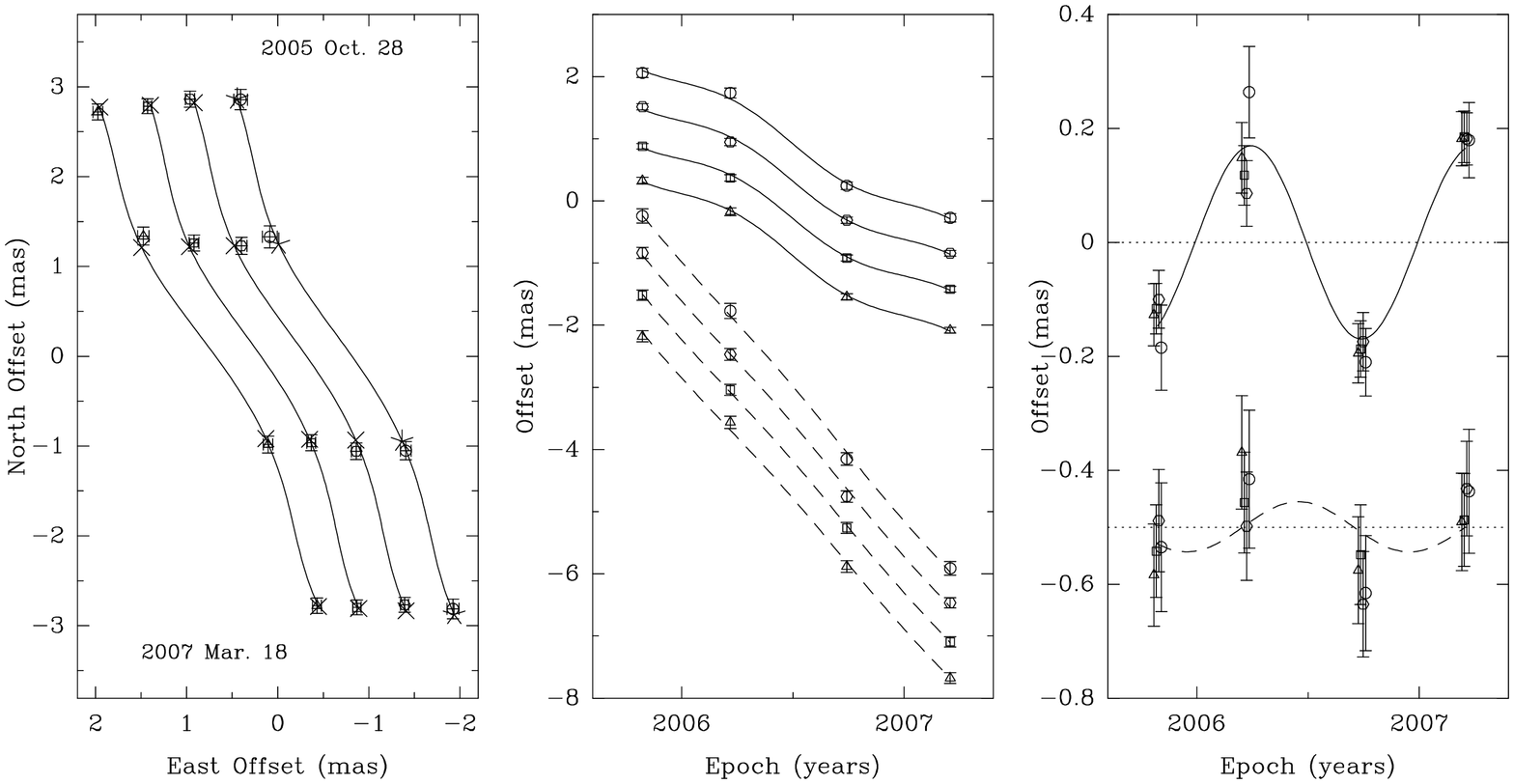}}
\caption{Results of the parallax fit for \tb. The different symbols represent 
the four maser spots. {\bf Left Panel:} Sky projected motion of the maser. The
crosses and the lines show the best-fit position offsets and the trajectrory,
respectively. {\bf Middle Panel:} The position offfsets of the masers along
the East and North direction versus time. The best-fit model in East and North
direction are shown as continuous and dashed lines, respectively. {\bf Right
  Panel:} Same as the middle panel bit with fitted proper motions removed. The
North offset data have been shifted for clarity.}
\label{g23.4}
\end{figure}

Toward \tb\ we detected several maser spots within a region of sky of
$\sim 50$~mas (see~Fig.~\ref{mom4}).  Four maser spots could be used for  
astrometric measurements.   Table~\ref{table:G4_ppm} and Fig.~\ref{g23.4} 
show the results of fitting for parallax and proper motion.   
The combined fit yields a parallax of 0.170 mas, with a formal
uncertainty of $\pm 0.016$~mas.  As for \ta, the post-fit (East) 
position residuals appear correlated, and we adopt the most conservative
approach of assuming 100\% correlation and multiplying the formal 
uncertainty by $\sqrt[]{4}$.  Hence, we obtain a distance of 
$5.88^{+1.37}_{-0.93}$~kpc for \tb. This translates to a 
Galactocentric radius of 3.9 kpc.

Our trigonometric parallax distance is in good agreement with the near 
kinematic distance of 5.6 kpc, but not the far distance of 10.0~kpc, 
assuming IAU values of $R_0=8.5$ and $\Theta_0=220$~\kms. 
Using R$_0$=8.0 kpc and $\Theta_0$=236 \kms\, would yield a smaller kinematic 
distance of 5.1 kpc.
The proper motions of the four maser spots are similar, and the 
average motion is $-1.93\pm 0.1$~\masy and $-4.11\pm 0.07$~\masy,
corresponding to $-54\pm10$~\kms\ and $115\pm22$\kms\ eastward and 
northward, respectively. 

\begin{deluxetable}{lllll}
\tablecolumns{5} \tablewidth{0pc}
\tablecaption{G23.44-0.18 Parallax \& Proper Motion Fits}
\tablehead {
  \colhead{Maser \Vlsr} & \colhead{Background} &
  \colhead{Parallax} & \colhead{$\mu_x$} &
  \colhead{$\mu_y$}
\\
  \colhead{(\kms)}      & \colhead{Source} &
  \colhead{(mas)} & \colhead{(\masy)} &
  \colhead{(\masy)}
            }
\startdata
98.4& J1825-0737 &$ 0.179\pm0.055$ &$ -1.97\pm0.10 $ &$ -4.05\pm0.09 $ \\
97.2& J1825-0737 &$ 0.159\pm0.028$ &$ -1.87\pm0.05 $ &$ -4.08\pm0.04 $ \\
96.9& J1825-0737 &$ 0.138\pm0.057$ &$ -1.88\pm0.10 $ &$ -4.12\pm0.09 $ \\
96.5& J1825-0737 &$ 0.230\pm0.046$ &$ -1.99\pm0.08 $ &$ -4.18\pm0.09 $ \\
\\
 combined &             &$0.170\pm0.032 $  \\

\enddata
\tablecomments {Combined fit used a single parallax parameter
for the four maser spots relative to the background source;
a single proper motion was fit for each maser spot relative
to the background source.}
\label{table:G4_ppm}
\end{deluxetable}

\section{Discussion}

Our measurements not only locate two sources in the Galaxy, 
they also provide the full space motion of each star-forming region.
A useful reference frame for Galactic dynamics removes a model
of Galactic rotation and yields peculiar motion relative to 
each source's ``local standard of rest.''  We define Cartesian
motion components U$_\mathrm{s}$,V$_\mathrm{s}$,W$_\mathrm{s}$ locally toward the Galactic center, 
toward direction of rotation, and toward the North Galactic Pole.
The methods used to convert the measured (heliocentric) proper motions 
into this frame will be described in detail in an upcoming paper.

If one adopts the IAU values for the LSR motion 
($R_0=8.5$ and $\Theta_0=220$ \kms) and the Hipparcos values for
the Solar Motin \citep{DehnenBinney1998}, 
we obtain the peculiar motion components listed in 
Table~\ref{table:galrotIAU}. Both masers move significant slower ($\sim$25 
\kms) than Galactic rotation and have large peculiar motions ($\sim$30 
\kms) toward the Galactic center.  
The peculiar motions toward the Galactic center could be reduced 
were one to use a rotation model of the Milky Way that is consistent 
with the measured proper motion of Sgr~A*, the super-massive
black hole at the Galactic center \citep{ReidBrunthaler2004}. 
However, changing the rotation model does not significantly
reduce the peculiar motions in the direction of Galactic rotation 
(Table~\ref{table:galrotRB}).

\begin{deluxetable}{llll}
\tablecolumns{4} \tablewidth{0pc}
\tablecaption{Peculiar Motions }  
\tablehead {
  \colhead{Maser}  & \colhead{U$_\mathrm{s}$} & 
  \colhead{V$_\mathrm{s}$}     & \colhead{W$_\mathrm{s}$} 
  \\
  \colhead{}       & \colhead{(\kms)} &
  \colhead{(\kms)} & \colhead{(\kms)}}

\startdata
\ta&37$\pm$6&$-$29$\pm$12&$-$1.4$\pm$3\\
\tb&23$\pm$12&$-$26$\pm$15&\phantom{$-$}1.9$\pm$3 \\
\enddata
\tablecomments {Peculiar motions assuming R$_0$=8.5, $\Theta_0$=220 \kms,
  and the Hipparcos Solar Motion \citep{DehnenBinney1998}. U$_\mathrm{s}$ is toward
  the Galactic center, V$_\mathrm{s}$ is toward Galactic rotation, and W$_\mathrm{s}$ is toward
  the North Galactic Pole.}
\label{table:galrotIAU}
\end{deluxetable}

\begin{deluxetable}{llll}
\tablecolumns{4} \tablewidth{0pc}
\tablecaption{Same as Table~\ref{table:galrotIAU} but for R$_0$=8.0 and $\Theta_0$=236 \kms}
\tablehead {
  \colhead{Maser}  & \colhead{U$_\mathrm{s}$} & 
  \colhead{V$_\mathrm{s}$}     & \colhead{W$_\mathrm{s}$} 
  \\
  \colhead{}       & \colhead{(\kms)} &
  \colhead{(\kms)} & \colhead{(\kms)}}

\startdata
\ta&\phantom{$-$}$21\pm 5$  &$-29\pm  8$ &$-1.4\pm 3$ \\
\tb&$-$5\phantom{-} $\pm 11$ &$-26\pm 13$ &\phantom{$-$}$1.9\pm 3$ \\
\enddata
\label{table:galrotRB}
\end{deluxetable}

\cite{BenjaminChurchwellBabler2005} find a stellar overdensity in the inner 
Galaxy, which can be explained with a central bar with half-length 
R$_\mathrm{bar}$=4.4 $\pm$ 0.5 kpc, tilted by $\Phi$=44$^\circ\pm10^\circ$ to 
the Sun-Galactic center line.   Other studies find smaller angles of 
$24^\circ-27^\circ$ between the major axis of the bar and the Sun-Galactic 
center and prefer a shorter bar \citep{Gerhard2002,RattenburyMaoSumi2007}.
Possibly there are multiple components to the bar.

\begin{figure}
\resizebox{\hsize}{!}{\includegraphics[angle=-90]{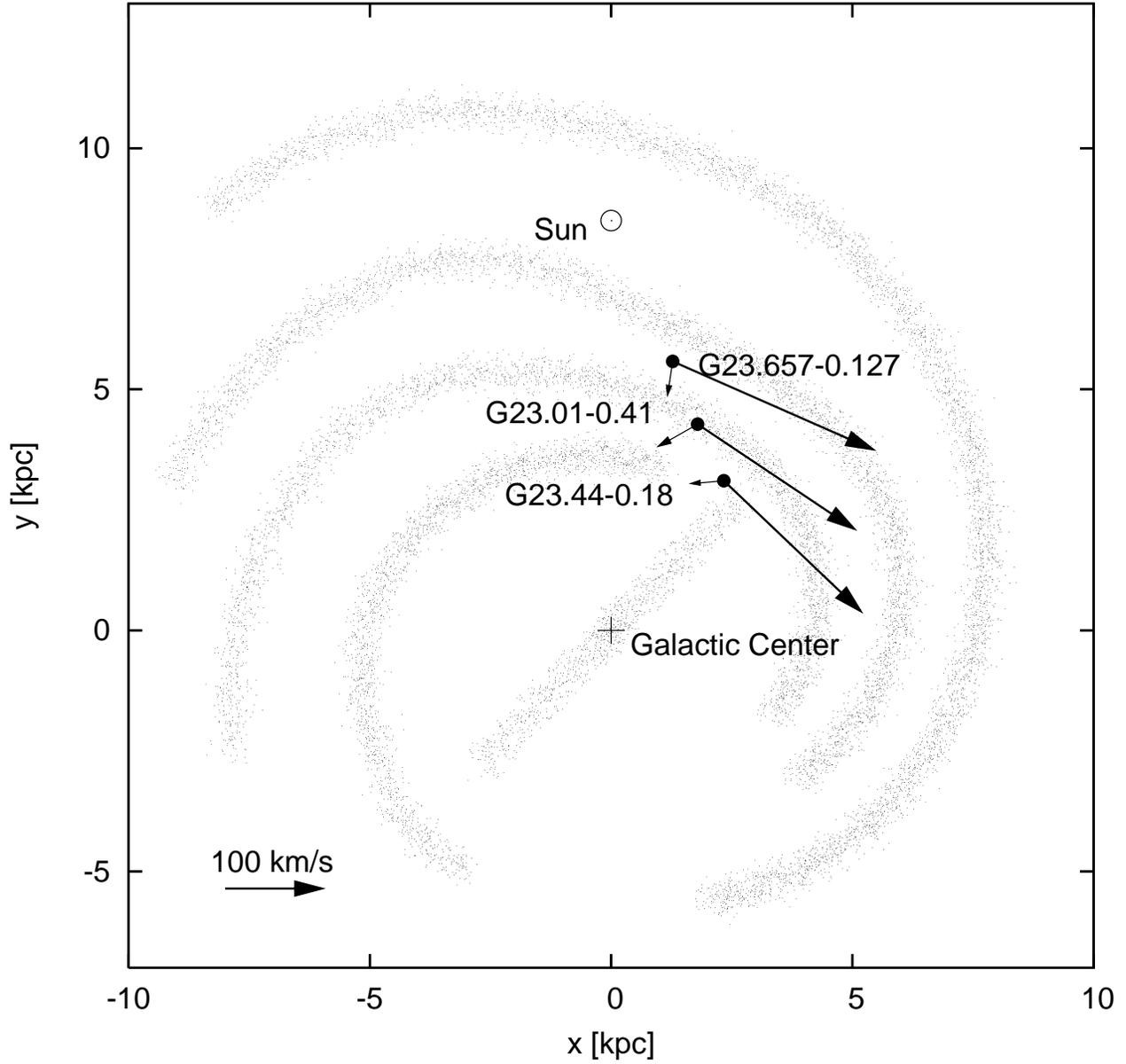}}
\caption{Schematic view of the Milky Way with the positions of \ta\, and \tb, their total motion relative to the Galactic Center (long arrows), and their peculiar motion after removing Galactic rotation (short arrows). The IAU values of R$_0$=8.5 kpc and $\Theta_o$=220 \kms\, are assumed. Also shown is the location of the central bar from \cite{BenjaminChurchwellBabler2005} and the position and motion of G23.657-0.127 by \cite{BartkiewiczBrunthalerSzymczak2008}.}
\label{mw}
\end{figure}

Our two star forming regions, in particular \tb\, are located close to the
end of the central bar (Fig.~\ref{mw}). This raises the question, whether 
the large peculiar motions of the sources are induced by the gravitational 
potential of the central bar. \cite{RobertsHuntleyvanAlbada1979} show that
a barlike potential can induce strong noncircular motions in the gas flow of 
$\sim50-150$ \kms.  Shocks may move gas in the inner parts outward and gas 
in the outer parts inward in a region, thus focusing gas where the spiral 
arm bends from the bar.  The space motions of our maser sources inward 
toward the Galactic center are qualitatively consistent with 
such bar shocks.

\acknowledgments
Andreas Brunthaler was supported by the DFG Priority Programme 1177. Ye Xu 
was supported by Chinese NSF through grants NSF 10673024, NSF 10733030, NSF 
10703010 and NSF 10621303.

\bibliography{brunthal_refs}

\begin{thebibliography}{}

\bibitem[\protect\citeauthoryear{{Araya} et~al.}{{Araya}
  et~al.}{2008}]{ArayaHofnerGoss2008}
{Araya} E.~D., {Hofner} P., {Goss} W.~M., et~al., 2008, ArXiv e-prints,
  0806.1548

\bibitem[\protect\citeauthoryear{{Bartkiewicz} et~al.}{{Bartkiewicz}
  et~al.}{2008}]{BartkiewiczBrunthalerSzymczak2008}
{Bartkiewicz} A., {Brunthaler} A., {Szymczak} M., {van Langevelde} H.~J.,
  {Reid} M.~J., 2008, \aap, 490, 787

\bibitem[\protect\citeauthoryear{{Benjamin} et~al.}{{Benjamin}
  et~al.}{2005}]{BenjaminChurchwellBabler2005}
{Benjamin} R.~A., {Churchwell} E., {Babler} B.~L., et~al., 2005, \apjl, 630,
  L149

\bibitem[\protect\citeauthoryear{{Codella}, {Testi}, \& {Cesaroni}}{{Codella}
  et~al.}{1997}]{CodellaTestiCesaroni1997}
{Codella} C., {Testi} L.,  {Cesaroni} R., 1997, \aap, 325, 282

\bibitem[\protect\citeauthoryear{{Dehnen} \& {Binney}}{{Dehnen} \&
  {Binney}}{1998}]{DehnenBinney1998}
{Dehnen} W.,  {Binney} J.~J., 1998, \mnras, 298, 387

\bibitem[\protect\citeauthoryear{{Fomalont} et~al.}{{Fomalont}
  et~al.}{2003}]{VCS2}
{Fomalont} E.~B., {Petrov} L., {MacMillan} D.~S., {Gordon} D.,  {Ma} C., 2003,
  \aj, 126, 2562

\bibitem[\protect\citeauthoryear{{Furuya} et~al.}{{Furuya}
  et~al.}{2008}]{FuruyaCesaroniTakahashi2008}
{Furuya} R.~S., {Cesaroni} R., {Takahashi} S., et~al., 2008, \apj, 673, 363

\bibitem[\protect\citeauthoryear{{Gerhard}}{{Gerhard}}{2002}]{Gerhard2002}
{Gerhard} O., 2002, {The Galactic Bar}, in Astronomical Society of the Pacific
  Conference Series, Vol. 273, {Da Costa} G.~S.,  {Jerjen} H. (eds.), The
  Dynamics, Structure \& History of Galaxies: A Workshop in Honour of Professor
  Ken Freeman, p.~73

\bibitem[\protect\citeauthoryear{{Ma} et~al.}{{Ma} et~al.}{1998}]{ICRF}
{Ma} C., {Arias} E.~F., {Eubanks} T.~M., et~al., 1998, \aj, 116, 516

\bibitem[\protect\citeauthoryear{{Moscadelli} et~al.}{{Moscadelli}
  et~al.}{2002}]{MoscadelliMentenWalmsley2002}
{Moscadelli} L., {Menten} K.~M., {Walmsley} C.~M.,  {Reid} M.~J., 2002, \apj,
  564, 813

\bibitem[\protect\citeauthoryear{{Petrov} et~al.}{{Petrov} et~al.}{2006}]{VCS4}
{Petrov} L., {Kovalev} Y.~Y., {Fomalont} E.~B.,  {Gordon} D., 2006, \aj, 131,
  1872

\bibitem[\protect\citeauthoryear{{Rattenbury} et~al.}{{Rattenbury}
  et~al.}{2007}]{RattenburyMaoSumi2007}
{Rattenbury} N.~J., {Mao} S., {Sumi} T.,  {Smith} M.~C., 2007, \mnras, 378,
  1064

\bibitem[\protect\citeauthoryear{{Reid} \& {Brunthaler}}{{Reid} \&
  {Brunthaler}}{2004}]{ReidBrunthaler2004}
{Reid} M.~J.,  {Brunthaler} A., 2004, \apj, 616, 872

\bibitem[\protect\citeauthoryear{{Reid} et~al.}{{Reid}
  et~al.}{2008}]{ReidMentenBrunthaler2008a}
{Reid} M.~J., {Menten} M.~K., {Brunthaler} A., et~al., 2008, \apj, submitted,
  arXiv:0811.0595

\bibitem[\protect\citeauthoryear{{Roberts}, {Huntley}, \& {van
  Albada}}{{Roberts} et~al.}{1979}]{RobertsHuntleyvanAlbada1979}
{Roberts} W.~W., Jr., {Huntley} J.~M.,  {van Albada} G.~D., 1979, \apj, 233, 67

\bibitem[\protect\citeauthoryear{{Xu} et~al.}{{Xu}
  et~al.}{2006}]{XuReidMenten2006}
{Xu} Y., {Reid} M.~J., {Menten} K.~M.,  {Zheng} X.~W., 2006, \apjs, 166, 526

\end{thebibliography}
\bibliographystyle{aa}

\end{document}